\documentclass[12pt]{article}
\usepackage{epsfig}

\newcommand\pubnumber{ER/40685/964 \\ UB-HET-01-01 \\ UR-1629}
\newcommand\pubdate{\today}
\newcommand\hepnumber{hep-ph/0101254}

\def\csumb{Department of Physics\\
State University of New York at Buffalo, Buffalo, NY 14260 USA}
\def\csumbb{Department of Physics and Astronomy\\
University of Rochester, Rochester, NY 14627 USA}
\def\support{\footnote{Work supported by NSF grants PHY-9970703 and
PHY-9600155, and by the U.S. Department of Energy, under grant
DE-FG02-91ER40685.}}

\def\Title#1{\begin{center} {\Large\bf #1 } \end{center}}
\def\Author#1{\begin{center}{ \sc #1} \end{center}}
\def\Address#1{\begin{center}{ \it #1} \end{center}}

\newcommand\pubblock{\rightline{\begin{tabular}{l} \pubnumber\\
         \pubdate\\ \hepnumber \end{tabular}}}
\newenvironment{Abstract}{\begin{quotation}  }{\end{quotation}}
\newenvironment{Presented}{\begin{quotation} \begin{center} 
             Presented at the\end{center}
      \begin{center}\begin{large}}{\end{large}\end{center} \end{quotation}}
\def\Acknowledgments{\bigskip  \bigskip \begin{center}
          \large\bf Acknowledgments\end{center}}

\makeatletter
\def\section{\@startsection{section}{0}{\z@}{5.5ex plus .5ex minus
 1.5ex}{2.3ex plus .2ex}{\large\bf}}
\def\subsection{\@startsection{subsection}{1}{\z@}{3.5ex plus .5ex minus
 1.5ex}{1.3ex plus .2ex}{\normalsize\bf}}
\def\subsubsection{\@startsection{subsubsection}{2}{\z@}{-3.5ex plus
-1ex minus  -.2ex}{2.3ex plus .2ex}{\normalsize\sl}}

\renewcommand{\@makecaption}[2]{%
   \vskip 10pt
   \setbox\@tempboxa\hbox{\small #1: #2}
   \ifdim \wd\@tempboxa >\hsize     
       \small #1: #2\par          
     \else                        
       \hbox to\hsize{\hfil\box\@tempboxa\hfil}
   \fi}

 \def\citenum#1{{\def\@cite##1##2{##1}\cite{#1}}}
 
\newcount\@tempcntc
\def\@citex[#1]#2{\if@filesw\immediate\write\@auxout{\string\citation{#2}}\fi
  \@tempcnta\z@\@tempcntb\m@ne\def\@citea{}\@cite{\@for\@citeb:=#2\do
    {\@ifundefined
       {b@\@citeb}{\@citeo\@tempcntb\m@ne\@citea\def\@citea{,}{\bf ?}\@warning
       {Citation `\@citeb' on page \thepage \space undefined}}%
    {\setbox\z@\hbox{\global\@tempcntc0\csname b@\@citeb\endcsname\relax}%
     \ifnum\@tempcntc=\z@ \@citeo\@tempcntb\m@ne
       \@citea\def\@citea{,}\hbox{\csname b@\@citeb\endcsname}%
     \else
      \advance\@tempcntb\@ne
      \ifnum\@tempcntb=\@tempcntc
      \else\advance\@tempcntb\m@ne\@citeo
      \@tempcnta\@tempcntc\@tempcntb\@tempcntc\fi\fi}}\@citeo}{#1}}
\def\@citeo{\ifnum\@tempcnta>\@tempcntb\else\@citea\def\@citea{,}%
  \ifnum\@tempcnta=\@tempcntb\the\@tempcnta\else
  {\advance\@tempcnta\@ne\ifnum\@tempcnta=\@tempcntb \else\def\@citea{--}\fi
    \advance\@tempcnta\m@ne\the\@tempcnta\@citea\the\@tempcntb}\fi\fi}
\makeatother

\def\beq{\begin{equation}}
\def\eeq#1{\label{#1}\end{equation}}
\def\eeqn{\end{equation}}

\newenvironment{Eqnarray}%
   {\arraycolsep 0.14em\begin{eqnarray}}{\end{eqnarray}}
\def\beqa{\begin{Eqnarray}}
\def\eeqa#1{\label{#1}\end{Eqnarray}}
\def\eeqan{\end{Eqnarray}}

\let\bar=\overbar

\def\O{{\cal O}}

\def\Dslash{\not{\hbox{\kern-4pt $D$}}}
\def\dslash{\not{\hbox{\kern-2pt $\del$}}}

\def\msb{{\bar{\ssstyle M \kern -1pt S}}}

\def\lsim{\mathrel{\raise.3ex\hbox{$<$\kern-.75em\lower1ex\hbox{$\sim$}}}}
\def\gsim{\mathrel{\raise.3ex\hbox{$>$\kern-.75em\lower1ex\hbox{$\sim$}}}}

\begin{document}
\begin{titlepage}
\pubblock

\vfill
\def\thefootnote{\fnsymbol{footnote}}
\Title{Theoretical Challenges for a Precision \\[5pt] 
Measurement of the W Mass at Hadron Colliders\support}
\vfill
\Author{Ulrich Baur}
\Address{\csumb}
\Author{Doreen Wackeroth}
\Address{\csumbb}
\vfill
\begin{Abstract}
We summarize the status of calculations of the electroweak radiative 
corrections to W and Z boson
production via the Drell-Yan mechanism at hadron colliders. To fully 
exploit the precision physics potential of the
high-luminosity environment of the Fermilab Tevatron $p \bar p$ (Run~II)
and the CERN 
LHC $pp$ colliders, it is crucial that the theoretical predictions are
well under control.  The envisioned precision physics program includes
a precise measurement of the W boson mass and the (leptonic) weak
mixing angle, as well as probing the Standard Model (SM) of
electroweak interactions at the highest accessible center-of-mass
energies. Some numerical results are presented. 
\end{Abstract}
\vfill
\begin{Presented}
5th International Symposium on Radiative Corrections \\ 
(RADCOR--2000) \\[4pt]
Carmel CA, USA, 11--15 September, 2000
\end{Presented}
\vfill
\end{titlepage}
\def\thefootnote{\arabic{footnote}}
\setcounter{footnote}{0}
\section{Introduction}

The Standard Model of electroweak interactions (SM) so far withstood
all experimental challenges and is tested as a quantum field theory at
the 0.1\% level~\cite{Abbaneo:2000nr}. However, the mechanism of mass
generation in the SM predicts the existence of a Higgs boson which, so
far, has eluded direct observation. Direct searches at LEP2 give 
a (preliminary) 95\% confidence-level lower bound on the mass of the
SM Higgs boson of $M_H> 113.5$~GeV~\cite{mhboundprem}.  Indirect
information on the mass of the Higgs boson can be extracted from the
$M_H$ dependence of radiative corrections to the W boson mass. With
the present knowledge of the W boson and top quark masses, and the
electromagnetic coupling constant, $\alpha(M_Z^2)$, the SM
Higgs boson mass can be indirectly constrained to 
$M_H=77^{+69}_{-39}$~GeV~\cite{Abbaneo:2000nr} by a global fit to all 
electroweak precision
data.  Future more precise measurements of the W boson and top quark
masses are expected to  considerably improve the present indirect bound on
$M_H$:~with a precision of 30~MeV for the W boson mass, $M_W$, and
2~GeV for the top quark mass which are target values for
Run~II of the Tevatron~\cite{Brock:1999ep}, 
$M_H$ can be predicted with an uncertainty
of about $30\%$. In addition, the confrontation of a precisely measured
W boson mass with the indirect SM prediction from a global fit to all
electroweak precision data, $M_W=80.385 \pm 
0.022$~GeV~\cite{Abbaneo:2000nr}, will provide a stringent test of
the SM. A detailed discussion of the prospects for the precision
measurement of $M_W$, and of the (leptonic) effective weak mixing
angle, $\sin^2\theta_{eff}^l$, at Run~II and the
LHC is given in Refs.~\cite{Brock:1999ep} and~\cite{Haywood:1999qg},
respectively.

In order to measure $M_W$ with high precision in a hadron collider
environment it is necessary to fully control higher order QCD and 
electroweak radiative corrections to the W and Z production processes.
The status of the QCD corrections to W and Z boson production 
at hadron colliders is reviewed in Refs.~\cite{Catani:2000jh,Baur:2000xd}. 
Here we discuss the electroweak ${\cal O}(\alpha)$ corrections to 
$p\,p\hskip-7pt\hbox{$^{^{(\!-\!)}}$} \to W^{\pm} \to l^{\pm} \nu_l$ and 
$p\,p\hskip-7pt\hbox{$^{^{(\!-\!)}}$} \to \gamma^*, Z \to l^+ l^-$ 
($l=e,\mu$) as
presented in detail in Refs.~\cite{Wackeroth:1997hz,Baur:1999kt} 
and~\cite{Baur:1998wa,Baurinprep}.

\section{{\boldmath Electroweak ${\cal O}(\alpha)$ Corrections to
$p\,p\hskip-7pt\hbox{$^{^{(\!-\!)}}$} \to W^{\pm} \to l^{\pm} \nu$}}

The full electroweak ${\cal O}(\alpha)$ corrections to resonant W
boson production in a general four-fermion process were
calculated in Ref.~\cite{Wackeroth:1997hz} with special emphasis on
obtaining a gauge invariant decomposition into a photonic and
non-photonic part. It was shown that the cross section for
resonant W boson production via the Drell-Yan mechanism at parton level, $q_i
\overline{q}_{i'}\rightarrow f \bar{f}'(\gamma)$, can be written in the
following form~\cite{Baur:1999kt}:
\begin{eqnarray}\label{eqwack:one}
d \hat{\sigma}^{(0+1)} & = &
d \hat \sigma^{(0)}\; [1+ 2 {\cal R}e (\tilde F_{weak}^{initial}(\hat s=M_W^2)+
\tilde F_{weak}^{final}(\hat s=M_W^2))] 
\nonumber \\
&+ & \sum_{a=initial,final,\atop interf.} [d\hat\sigma^{(0)}\; 
F_{Q\!E\!D}^a(\hat s,\hat t)+
d \hat \sigma_{2\rightarrow 3}^a] \; ,
\end{eqnarray}
where the Born cross section, $d \hat \sigma^{(0)}$, is of
Breit-Wigner form, and $\hat s$ and $\hat t$ are the usual Mandelstam
variables in the parton center of mass frame.  The (modified) weak
corrections and the virtual and soft photon emission from the initial
and final state fermions (as well as their interference) are described
by the form factors $\tilde F_{weak}^a$ and $F_{Q\!E\!D}^a$,
respectively.  The IR finite contribution $d\hat\sigma_{2\rightarrow
3}^a$ describes real photon radiation away from soft singularities.
Mass singularities of the form $\ln(\hat s/m_f^2)$ arise 
when the photon is emitted collinear with  
a charged fermion and the resulting singularity is regularized 
by retaining a finite fermion mass ($m_f$).
$F_{Q\!E\!D}^{initial}$ and $d\hat\sigma_{2\rightarrow 3}^{initial}$
still include quark-mass singularities which need to be extracted and
absorbed into the parton distribution functions (PDFs).  
The absorption of the quark-mass singularities into the PDFs can be
done in complete analogy to gluon emission in QCD, thereby introducing
a QED factorization scheme dependence.  Explicit expressions for the W
production cross section in the QED DIS and $\overline{\mathrm{MS}}$
scheme are provided in Ref.~\cite{Baur:1999kt}.  So far, in the
extraction of the PDFs from data as well as in the PDF evolution, QED
corrections are not taken into account.  The latter result in a
modified scale dependence of the PDFs, which is expected to have a
negligible effect on the observable cross
sections~\cite{Haywood:1999qg}.  The numerical evaluation of the cross
section is done with the parton level Monte Carlo program
{\tt WGRAD}~\cite{Baur:1999kt}\footnote{{\tt WGRAD} is available from the
authors.}, and results have been obtained for a variety of
interesting W boson observables at the Tevatron~\cite{Baur:1999kt} and
the LHC~\cite{Haywood:1999qg}.  

\begin{figure}[htb]
\begin{center}
\epsfig{file=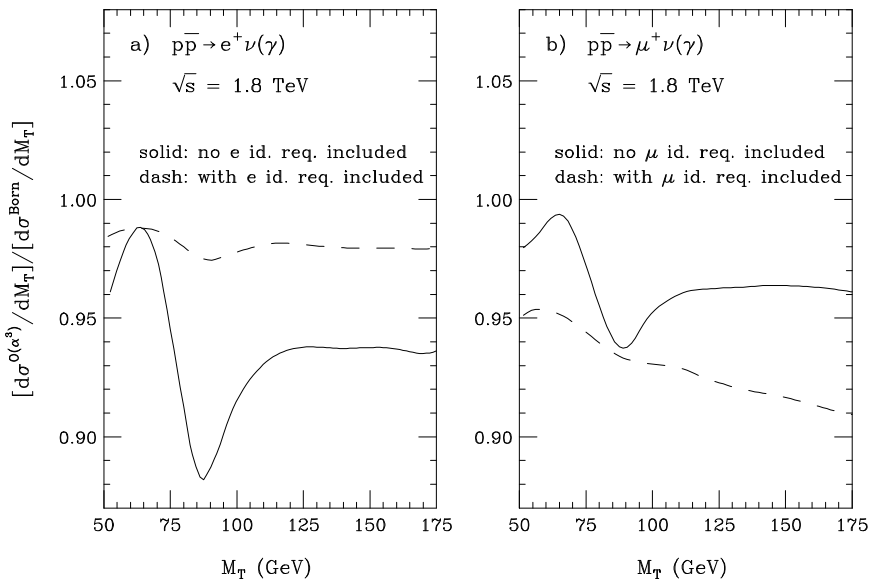,height=3in}
\caption[0]{The relative corrections to the $M_T(l \nu)$ distributions
at the Tevatron when taking into account the full electroweak
${\cal O}(\alpha)$ corrections
(from Ref.~\cite{Baur:1999kt}).}\label{fig:one}
\end{center}
\end{figure}

In the past, fits to the distribution
of the transverse mass of the final-state lepton neutrino system,
$M_T(l\nu)$, have provided the most accurate measurements of $M_W$~\cite{mw}.
Photonic initial state and initial-final state interference corrections were
found to have only a small effect on the $M_T$ distribution, and weak
corrections uniformly reduce the cross section by about $1 \%$.
However, final-state photon
radiation significantly distorts the shape of the $M_T$ distribution,
and thus considerably affects the extracted value of $M_W$.  In the
electron case, when taking into account realistic lepton
identification requirements to simulate the detector acceptance, 
the electroweak radiative corrections are
strongly reduced because electron and photon momenta are combined for
small opening angles between the two particles. This eliminates the
mass singular terms associated with final state radiation. The ratio of
the full ${\cal O}(\alpha^3)$ and lowest order differential cross
section as a function of $M_T(l\nu)$ with and without lepton
identification requirements taken into account is shown in
Fig.~\ref{fig:one}. 

A  previous approximate calculation~\cite{Berends:1985qa} took only the
real photonic corrections properly into account while the effect of soft
and virtual virtual photonic corrections were estimated from the
inclusive $W\to l\nu(\gamma)$ width. Weak corrections were ignored in 
Ref.~\cite{Berends:1985qa}. Comparing the W mass shifts obtained using
the calculations of Refs.~\cite{Berends:1985qa} and~\cite{Baur:1999kt},
one finds that the proper treatment of virtual and soft corrections and
the inclusion of weak corrections induces an additional shift of ${\cal
O}$(10 MeV) in the extracted W boson mass.

\section{{\boldmath Electroweak ${\cal O}(\alpha)$ Corrections to
$p\,p\hskip-7pt\hbox{$^{^{(\!-\!)}}$} \to\gamma^*,Z \to l^+ l^-$}}

Neutral-current Drell-Yan production is interesting for several
reasons:
\begin{enumerate}

\item 
Future precise measurements of the W
boson mass at hadron colliders depend on a precise knowledge of the Z
boson production process. When compared to the values measured at LEP,
the measured Z boson mass and width help to determine the energy scale
and resolution of the electromagnetic calorimeter. 

\item Ratios of W and Z
boson observables may yield a more precise measurement of $M_W$ than
the traditional technique of fitting the $M_T$
distribution~\cite{Brock:1999ep,gk}.  

\item
The forward-backward asymmetry in
the vicinity of the Z resonance can be used to measure the (leptonic) effective
weak mixing angle~\cite{Haywood:1999qg,Baur:1998wa}. Studying the 
forward-backward
asymmetry above the Z resonance probes the $\gamma,Z$ interference at
the highest available energies. 

\item
Finally, at large di-lepton
invariant masses, $m(l^+l^-)$, deviations from the SM prediction could
indicate the presence of new physics, such as new heavy gauge bosons
$Z'$ or extra spatial dimensions.
\end{enumerate}
It is therefore important to determine the electroweak
corrections for this process. 

The electroweak ${\cal O}(\alpha)$ corrections to neutral-current
Drell-Yan processes naturally decompose into QED and weak
contributions, i.e.~they build gauge invariant subsets and thus can
be discussed separately.  The observable next-to-leading order (NLO) 
cross section is obtained
by convoluting the parton cross section with the quark distribution
functions $q(x,Q^2)$ ($\hat s=x_1 x_2 S$)~\cite{Baurinprep}
\begin{equation}\label{eq:xsecobs}
{\rm d}\sigma(S)  =  \int_0^1 d x_1 d x_2 \, q(x_1,Q^2) \, \bar q(x_2,Q^2) 
[{\rm d}\hat \sigma^{(0+1)}(\hat s,\hat t)+ 
{\rm d}\hat \sigma_{{\rm QED}}(\mu_{{\rm QED}}^2,\hat s,\hat t)] \; ,   
\end{equation}
where ${\rm d}\hat \sigma^{(0+1)}$ comprises the NLO cross section
including weak corrections, and ${\rm d}\hat \sigma_{{\rm QED}}$
describes the QED part, i.e.~virtual corrections and real photon 
emission off the
quarks and charged leptons. The PDFs depend on the QCD renormalization
and factorization scales which we choose to be equal; the common scale
is denoted by
$Q^2$.  The radiation of collinear photons off quarks requires the
factorization of the arising mass singularities into the PDFs which
introduces a dependence on the QED factorization scale, $\mu_{QED}$. 
The treatment of mass singularities is universal and
thus the same as in the $W$ case. The QED ${\cal O}(\alpha)$ corrections to
$p\,p\hskip-7pt\hbox{$^{^{(\!-\!)}}$} \to\gamma^*,Z \to l^+ l^-$
($l=e,\mu$) have been calculated and implemented in the parton level
Monte Carlo program {\tt ZGRAD}~\cite{Baur:1998wa}\footnote{{\tt ZGRAD} is
available from the authors.} and their impact 
on the di-lepton invariant mass spectrum, the 
lepton transverse momentum distribution, and  
on the forward-backward asymmetry, $A_{{\rm FB}}$, 
has been studied. In addition, the
prospects for a precision measurement of $\sin^2\theta_{eff}^l$ 
extracted from $A_{{\rm FB}}$ at the Z resonance at the LHC 
were investigated.

\begin{figure}[htb]
\begin{center}
\epsfig{file=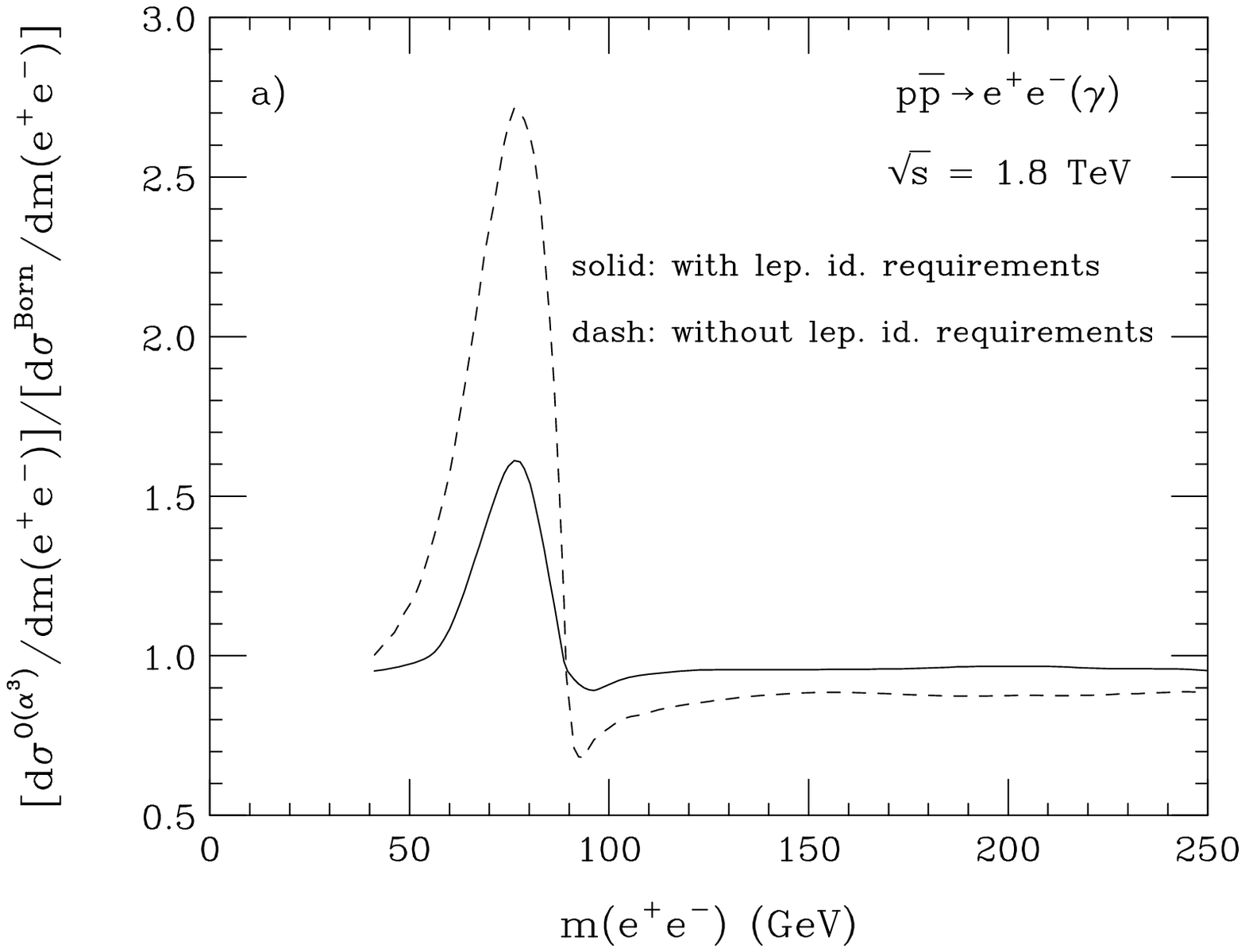,height=2.25in}
\epsfig{file=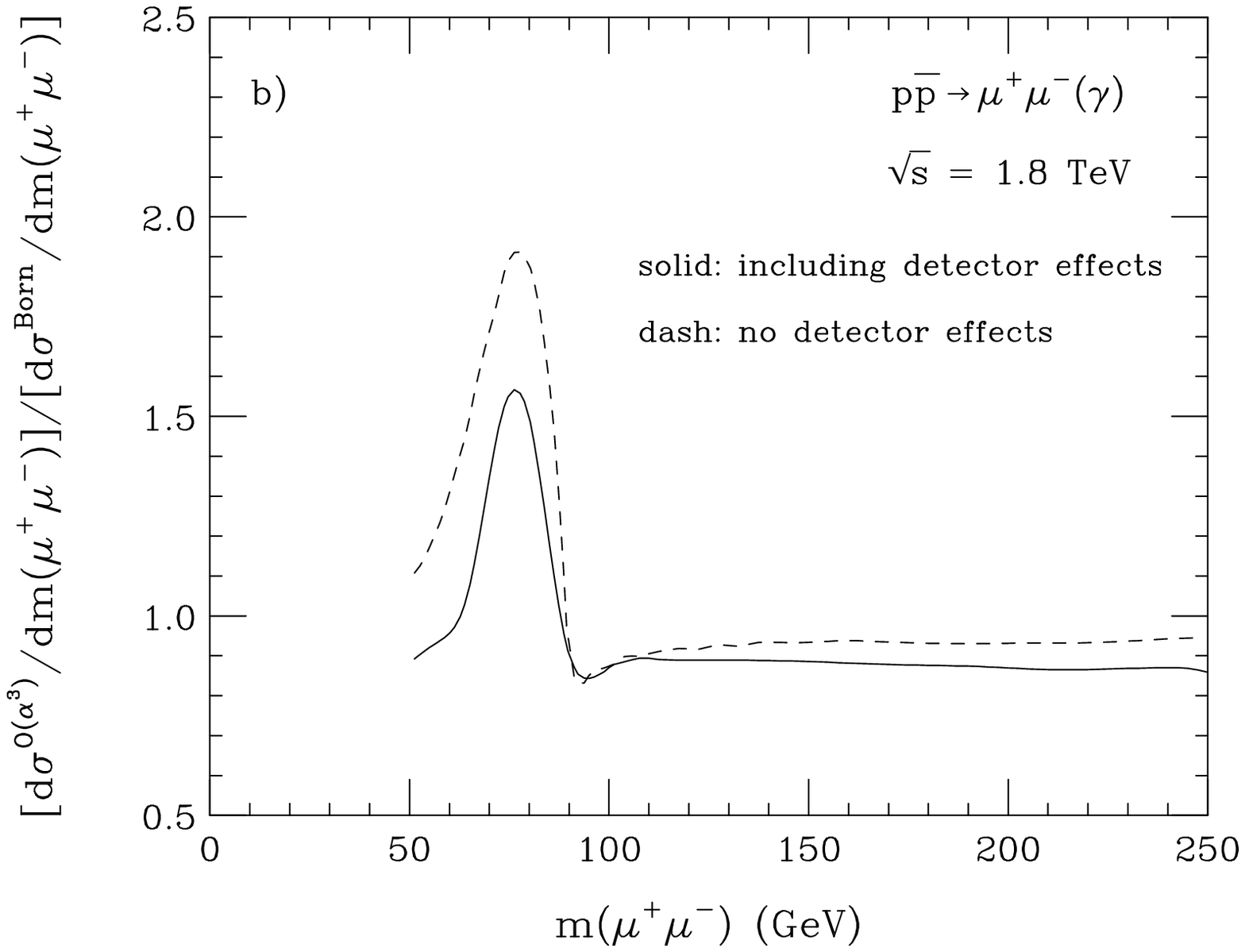,height=2.25in}
\caption[0]{The relative corrections to the $m(e^+ e^-)$ and
$m(\mu^+ \mu^-)$ distributions in Drell-Yan production at the Tevatron 
due to the ${\cal O}(\alpha)$ QED corrections 
(from Ref.~\cite{Baur:1998wa}). }\label{fig:two}
\end{center}
\end{figure}

In Fig.~\ref{fig:two} we show the effect of the QED corrections on
the invariant mass distribution of the final state lepton pair.
Similar to the transverse mass distribution in the charged-current 
Drell-Yan process,
final-state photon radiation strongly affects the shape of the $m(l^+l^-)$
distribution. When lepton identification requirements are taken into
account, the large contributions from mass singular logarithms
largely cancel in the electron case. As in the charged-current Drell-Yan
process, initial-final state interference is negligible, and the impact of
initial-state radiation is small after factorizing the collinear
singularities into the PDFs.
The difference in the extracted Z boson mass when comparing 
the approximate calculation of Ref.~\cite{Berends:1985qa}
with the full calculation of the ${\cal O}(\alpha)$ QED corrections
is of ${\cal O}$(10 MeV). Since the detector response is calibrated
using Z boson observables, the shift in the Z boson mass is expected 
to slightly modify the W mass extracted from experiment.

For precision physics away from the Z resonance, the (non-universal) weak
corrections must also be included.  These corrections 
become important at large values of the di-lepton invariant mass 
due to the presence of large
Sudakov-like electroweak logarithms of the form $\ln(m(l^+l^-)/M_V)$,
$V=W,\,Z$, which eventually may be resummed~\cite{resummation}. A 
calculation of the non-universal weak
corrections in $p\,p\hskip-7pt\hbox{$^{^{(\!-\!)}}$} \to\gamma^*,Z \to
l^+ l^-$ is currently in progress~\cite{Baurinprep}. In the implementation
of the weak corrections we closely follow Ref.~\cite{yellowbook95},
in particular for the treatment of higher-order
corrections, which are important for a precise description of the Z
resonance.  

The NLO parton differential cross section, including weak 
${\cal O}(\alpha)$ and leading ${\cal O}(\alpha^2)$ corrections, which enters 
eq.~(\ref{eq:xsecobs}) is of the form~\cite{Baurinprep} 
\begin{equation}\label{eq:xsecweak}
{\rm d} \hat \sigma^{(0+1)}={\rm dP_{2f}} \, \frac{1}{12} \, \sum 
|A_{\gamma}^{(0+1)}+ A_Z^{(0+1)}|^2(\hat s,\hat t) + 
{\rm d} \hat \sigma_{{\rm box}}(\hat s,\hat t) \; ,
\end{equation} 
where the sum is taken over the spin and color degrees of freedom, and 
${\rm dP_{2f}}$ denotes the two-particle phase space.
${\rm d} \hat \sigma_{{\rm box}}$ describes the contribution of the
box diagrams involving two massive gauge bosons.  The matrix elements
$A_{\gamma,Z}^{(0+1)}$ comprise the Born matrix elements, the
$\gamma,Z, \gamma Z$ self energy insertions including a leading-log
resummation of the terms involving the light fermions, and the
one-loop vertex corrections. While $A_{\gamma,Z}^{(0+1)}$ can be expressed
in terms of effective vector and axial-vector couplings,  the box
contribution ${\rm d}\hat \sigma_{{\rm box}}$ cannot be absorbed in
effective couplings. However, in the Z resonance region the box
diagrams can be neglected and the NLO cross section ${\rm d}\hat
\sigma^{(0+1)}$ of eq.~(\ref{eq:xsecweak}) has a Born-like structure.
The leading universal electroweak corrections, 
i.e.~the running of the electromagnetic charge and 
corrections connected to $\Delta \rho$, can be included in form of an 
effective Born approximation (EBA). Comparing results of the
calculation which includes the full ${\cal O}(\alpha)$ corrections with
those obtained using the EBA together with the pure QED corrections 
reveals the effects of the genuine
non-universal electroweak corrections such as box diagrams. 

The weak corrections to neutral-current Drell-Yan processes as
described above are currently being implemented in the parton level MC program
{\tt ZGRAD2}~\cite{Baurinprep}.  A detailed numerical discussion 
of the effects of the electroweak ${\cal O}(\alpha)$ corrections on
distributions in $p\,p\hskip-7pt\hbox{$^{^{(\!-\!)}}$} \to \gamma^*,Z
\to l^+l^- (\gamma), l=e,\mu$ at the Tevatron and the LHC will be given in
Ref.~\cite{Baurinprep}.  Here we present some selected preliminary results
for the di-lepton invariant mass distribution and the forward-backward
asymmetry. 
 
\begin{figure}[tb]
\begin{center}
\epsfig{file=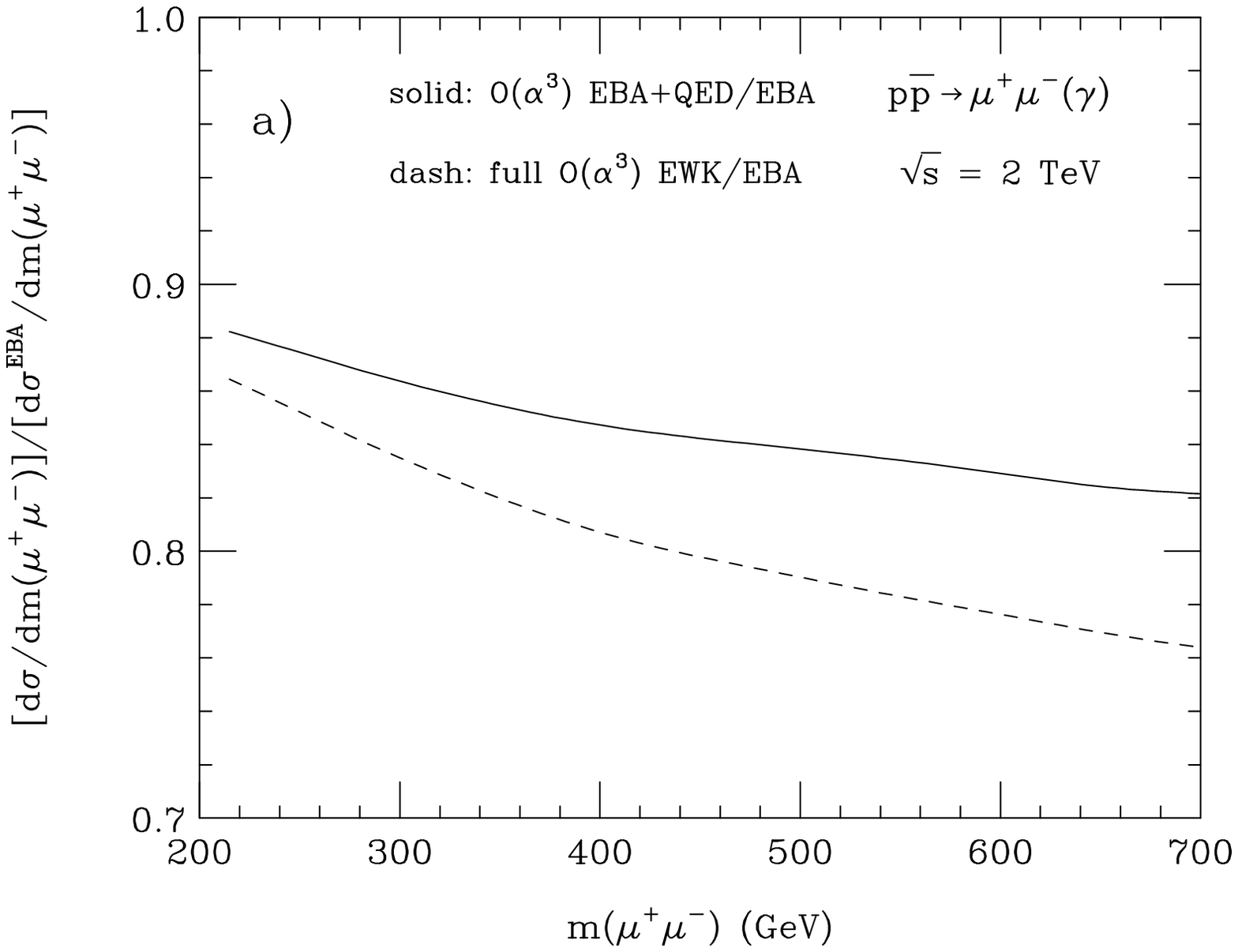,height=2.25in}
\epsfig{file=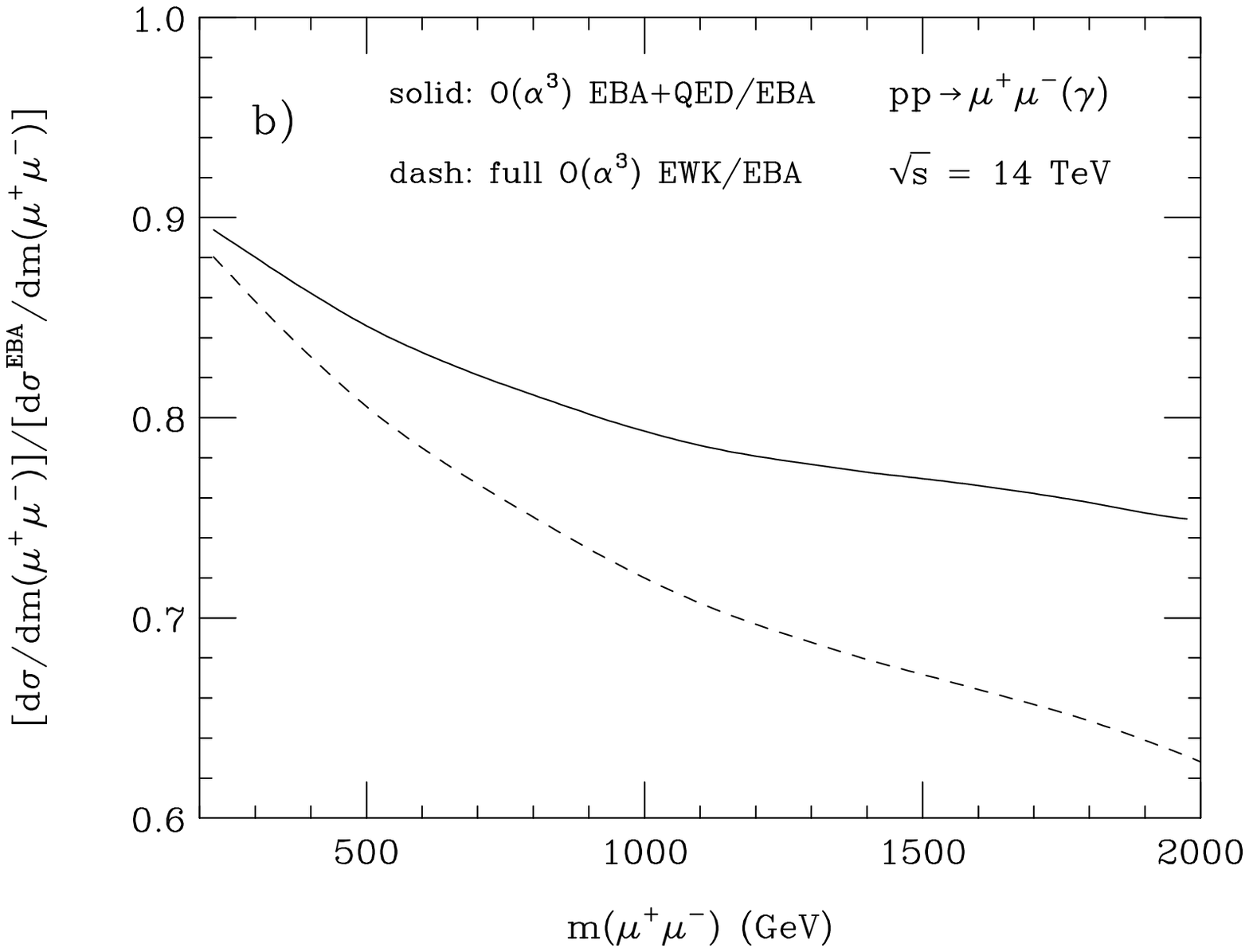,height=2.25in}
\caption[0]{The relative corrections to the $m(\mu^+ \mu^-)$ distribution
a) at the Tevatron and b) at the LHC when taking into account the universal
corrections entering the EBA and 
QED corrections only (solid line), and when the full ${\cal O}(\alpha)$
electroweak corrections are included in the calculation 
(dashed line).}\label{fig:three}
\end{center}
\end{figure}

In Fig.~\ref{fig:three} we show the $\mu^+ \mu^-$ invariant mass
distribution including the full ${\cal O}(\alpha)$ corrections
normalized to the differential cross section in the EBA for
large di-lepton invariant masses at
the Tevatron and the LHC. Separation
cuts and lepton identification requirements to simulate the detector
acceptance as described in Ref.~\cite{Baur:1998wa} (Tevatron) and
Ref.~\cite{Haywood:1999qg} (LHC) are taken into account in 
Fig.~\ref{fig:three}.  For comparison the relative
corrections including the QED corrections only are also shown.  
As expected from the presence of large
electroweak Sudakov-like logarithms, the weak corrections strongly
increase in magnitude with increasing $m(\mu^+ \mu^-)$, reaching about
10\% at $m(\mu^+ \mu^-)=1$~TeV. Both, the QED and the genuine weak
corrections reduce the differential cross section.
Qualitatively similar results are
obtained in the $e^+e^-$ case.

\begin{figure}[t]
\begin{center}
\epsfig{file=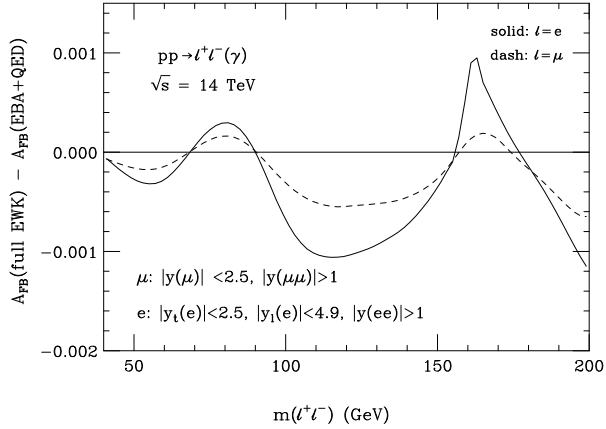,height=2.25in}
\caption[0]{The forward-backward asymmetry including NLO electroweak
corrections at the LHC, imposing 
the cuts and lepton identification requirements of Ref.~\cite{atlas}. 
The EBA and QED contribution have been subtracted 
(preliminary results).}\label{fig:four}
\end{center}
\end{figure}

In Fig.~\ref{fig:four}, we show how the purely weak corrections affect
the forward backward asymmetry at the LHC\footnote{For a definition of 
$A_{{\rm FB}}$ at the LHC, see Ref.~\cite{Baur:1998wa}.}. To
illustrate the effect of the non-universal weak corrections, we plot the
difference of the forward backward asymmetry including the full ${\cal
O}(\alpha)$ corrections, and the asymmetry which only takes into account
QED corrections and the universal corrections which are included in the
EBA. A genuine non-universal electroweak effect can be
observed in the vicinity of $m(l^+l^-)=M_W$ and $2\,M_W$, which is due
to threshold effects in the box diagrams involving two W
bosons. Results qualitatively similar to those shown in
Fig.~\ref{fig:four} are also obtained for the Tevatron. 

The forward backward asymmetry at the LHC is very sensitive to the
rapidity coverage of the leptons assumed. In Fig.~\ref{fig:four}, we
have used the lepton rapidity coverages foreseen for the ATLAS
detector~\cite{Haywood:1999qg,atlas}. For muon pairs, both muons are
required to have rapidity $|y(\mu)|<2.5$. For $e^+e^-$ pairs, the
leptons are required to have $|y_l(e)|<4.9$, with one of them having 
to fulfill the more stringent requirement $|y_t(e)|<2.5$. In addition,
the lepton pair rapidity has to be $|y(ll)|>1$ for both electrons and
muons in the final state. This cut substantially increases the magnitude
of $A_{{\rm FB}}$ at the LHC~\cite{dittmar}. 

It is interesting to check whether the threshold effect at
$m(l^+l^-)=2\,M_W$ will be observable.
In the electron case, the expected statistical uncertainty in $A_{{\rm FB}}$
for $m(e^+e^-)=2\,M_W\pm 5$~GeV and 100~fb$^{-1}$ at the LHC is about 
$(3-4)\times 10^{-3}$ per experiment. The size of the non-universal
electroweak corrections in the region are 
of the order of $10^{-3}$. In a realistic calculation, contributions
from $W^+W^-\to l^+\nu_l l^-\bar\nu_l$, $ZZ\to l^+l^-\bar\nu\nu$ and
$\bar tt$ production to the forward backward asymmetry need to be
taken into account, which could well be of the same order as the genuine
weak corrections. It will thus be difficult to observe a clear signal of
the threshold effects originating from the box diagrams involving two W
bosons in $A_{{\rm FB}}$ at the LHC. On the other hand, given the expected
statistical precision, the genuine weak corrections cannot be neglected
when comparing data with the SM prediction.

\section{Conclusions}

Our results show that, for the precision obtained in previous Tevatron 
runs, the existing calculations for W and Z boson production are
sufficient. However, for future precision measurements the full
electroweak ${\cal O}(\alpha)$ corrections and probably also multiple photon
radiation effects should be taken into account.  The inclusion of the
non-resonant contributions to W production in {\tt WGRAD} is in
progress~\cite{wbinprep} (see also Ref.~\cite{dkinprep}).  
As a first step towards a calculation of 
the ${\cal O}(\alpha^2)$ QED corrections, the effects of
two-photon radiation in W and Z boson production at
hadron colliders have been computed in Ref.~\cite{Baur:2000hm}.

\Acknowledgments
We would like to thank the organizers of RADCOR-2000, especially H.~Haber,
for a delightful and inspiring conference experience, and for
arranging the spectacular sunsets at Carmel beach. One of us (U.B.) is 
grateful to the Fermilab Theory Group,
where part of this work was carried out, for its generous hospitality.

\end{document}